\documentstyle[prl,aps]{revtex}

\begin{document}

\title{Solid molecular hydrogen: The Broken Symmetry Phase} 
\author{Jorge Kohanoff~$^{1}$, Sandro Scandolo~$^{1,2,3}$, Guido L.
Chiarotti~$^{2,3}$ and Erio Tosatti~$^{1,2,3}$}
\address{
$^{1)}$ International Centre for Theoretical Physics, \\
Strada Costiera 11, I-34014 Trieste, Italy \\
$^{2)}$ International School for Advanced Studies (SISSA) \\
via Beirut 4, I-34014 Trieste, Italy\\
$^{3)}$ Istituto Nazionale Fisica della Materia (INFM) }
\date{\today}
\maketitle

\begin{abstract}
By performing constant-pressure variable-cell {\it ab initio} molecular 
dynamics simulations we find a quadrupolar orthorhombic structure, 
of $Pca2_1$ symmetry, for the broken symmetry phase (phase II) of 
solid H$_2$ at $T=0$ and P $=110 - 150$ GPa. We present results for the
equation of state, lattice parameters and vibronic frequencies, in very
good agreement with experimental observations. Anharmonic quantum corrections
to the vibrational frequencies are estimated using available
data on H$_2$ and D$_2$. We assign the observed modes to specific
symmetry representations.
\end{abstract}
\newpage

\twocolumn

The quest for the structure of the high-pressure phases of hydrogen
is a long standing one. Early predictions of an insulator-metal 
transition~\cite{wigner}, led to a large body of experimental and 
theoretical work during the past sixty years. Metallization 
was not found as promptly as initially expected~\cite{nellis} but
a rich phase diagram emerged. The current picture 
of the low and room-temperature phase diagram up to $\sim 230$ GPa is 
essentially based on the optical studies performed in diamond anvil cell (DAC)
devices during the past decade~\cite{hemley1,silvera1}.

There is a consensus that hydrogen exhibits at least three different solid 
molecular phases. 1) At low pressures ($< 110$ GPa for para-H$_2$) the 
centers of the H$_2$ molecules crystallize into an {\it hcp} lattice, but 
zero-point motion overcomes rotational energy barriers,
leading to a free-rotator phase (phase I). 2) Between 110 and 150 GPa 
intermolecular interactions freeze the molecular rotations into an 
ordered broken-symmetry phase (BSP, or phase II). 
3) Above $\sim 150$ GPa, a third phase 
(H-A, or phase III) is attained, whose structure is unknown.
Here we focus specifically on phase II. 
Optical measurements in phase II indicate the presence of
two, possibly three, infrared (IR) active modes, in constrast to
phase I, where only one mode is observed. Raman spectra show a
single peak, at a frequency $\sim 10$ \% lower than that of the IR
modes. A consistent picture of the structural and dynamical 
properties of phase II is still lacking~\cite{silvera2}.

On the theoretical side, most of the existing work consists of 
static total energy calculations within the local density 
approximation (LDA). Zero-point energy (ZPE) of the protons has been 
in a few cases included {\it a posteriori} based on frozen phonon 
calculations~\cite{barbee,surh}. 
Hexagonal close packed structures with two and four molecules 
per unit cell appear to be the strongest candidates for the ground structure 
of phase II, but the relative 
orientation of the molecules is uncertain~\cite{kaxiras,nagara,mazin}. 
Cubic structures were also suggested~\cite{nagara}. This uncertainty 
persists, due to the incomplete optimization of the lattice parameters and
the {\em a priori} selection of a particular space group symmetry in static 
calculations.
Ab initio Molecular Dynamics simulations, 
which do not rely on the choice of a specific space group, have also 
been reported~\cite{hohl,klug}. However, these earlier attempts were hampered 
by a fixed simulation cell and a poor Brillouin Zone (BZ) sampling.

We report extensive ab initio Molecular Dynamics simulations of 
solid H$_2$ in the $110 - 150$ GPa range. Additional crucial ingredients
are ($i$) a variable cell, constant pressure approach 
that largely eliminates prejudice on the space group symmetry, 
molecular orientations and lattice constants~\cite{focher}; 
($ii$) fully converged BZ sampling, achieved by a freshly implemented 
$k\cdot p$ technique~\cite{scandolo}. 
Electronic correlations are treated within the LDA supplemented with 
Becke-Perdew gradient corrections~\cite{beckper}, while the proton-electron 
interaction is described through a local pseudopotential~\cite{giannozzi} 
requiring an energy cutoff of 60 Ry~\cite{note}. We stress that 
electrostatic and band energies are fully included in this 
approach~\cite{mazin}. 

Preliminary runs using large simulation cells (up to 128 atoms) with 
$\Gamma$-point sampling only, produced layered ground state structures 
with in-plane triangular ordering of molecules. However, we found 
a strong dependence of the results on cell shape and atom number,
confirming that an accurate BZ sampling is crucial~\cite{mazin,ashcroft}. 
Convergence on BZ sums for insulating molecular H$_2$ turned out to require
at least $8\times 8\times 8$ $k$-points in the full BZ of 
a four-molecule cell~\cite{mazin}. This is 
particulary demanding in our approach because of the lack of 
symmetry constraints. Thus, at variance with the standard
approach where the electronic orbitals for each $k$-point in the BZ 
are expanded on a generic basis set (e.g. plane waves), we adopt a $k\cdot p$
expansion in terms of the (occupied plus empty) self-consistently 
generated $\Gamma$-point orbitals:

\begin{equation}
u_i^{\bf k}({\bf r})=\sum_{i=1}^{N}~a_{ij}^{\bf k}~u_j^{\bf 0}({\bf r})~~,
\end{equation}

\noindent where $u_i^{\bf k}({\bf r})$ is the periodic part of the
$i$-th Bloch function at wave number ${\bf k}$, and $a_{ij}^{\bf k}$
are unitary matrices obtained by direct diagonalization the $k\cdot p$ 
hamiltonian at ${\bf k}$~\cite{harrison}. This expansion is strictly equivalent
to the standard approach when the number of states $N$ equals the 
size of the basis set~\cite{payne}. 
In practice the sum can be truncated at a significantly 
smaller number of $\Gamma$-point orbitals with excellent 
accuracy~\cite{scandolo}. 
With this improved functional we studied 
a system of 8 atoms with 512 $k$-points in the full BZ~\cite{monkhorst}, 
and with $N=64$.
We also performed simulations with 32 atoms, 128 $k$-points and 
$N=96$, and with 64 atoms, 64 $k$-points and $N=128$. 

The 8-atom simulation started from the $Pmc2_1$ structure 
proposed by Kaxiras and Broughton~\cite{kaxiras} (see Fig. 1), 
that is the most favorable among a class of structures 
containing also the $P2/m$. Pressure was set to 140 GPa. 
When constant-pressure dynamics was switched on, the system spontaneously 
transformed into Kitaigorodskii and Mirskaya's 
quadrupolar structure $Pca2_1$~\cite{KM}, proposed for H$_2$ by Nagara 
and Nakamura~\cite{nagara} (see Fig. 1). We then performed additional runs 
starting with different initial structures, namely the cubic $Pa3$ and 
the orthorhombic $Cmca$ (see Fig. 1), previously  
proposed for solid H$_2$ at lower and higher pressures, 
respectively~\cite{silvera1,ashcroft}. Although these structures can be
be ruled out for phase II on the basis of the number of vibronic 
modes~\cite{silvera2}, they represent reasonable starting points 
for our dynamical structural search. In the range of existence of phase II,
the $Cmca$ structure turned out to be unstable against $Pca2_1$, 
while $Pa3$ was found to be locally stable, but much higher in enthalpy
than $Pca2_1$. The quadrupolar structure
of symmetry $P2_1/c$~\cite{nagara} was also considered and found to be
locally stable, with an enthalpy 0.015 eV/molecule higher 
than that of the $Pca2_1$ (at 140 GPa). 
The simulations were repeated with supercells of 32 atoms 
(double size along the two
in-plane directions), and 64 atoms (double size in all directions). 
However, we failed to observe any signature of ground state structures with 
unit cells larger than 8 atoms, and $Pca2_1$ prevailed in all cases. 

The above results have been obtained by treating the protons as classical
particles, since quantum effects have been proven irrelevant to the relative 
hierarchy of molecular phases~\cite{natoli}. However, given the high level 
of accuracy of the present calculations, it seems wise to include the 
quantum effects. By rotating a single molecule on the $Pca2_1$ 
structure~\cite{h16} we do not find multiple energy wells, suggesting that 
tunnelling (of high order in $\hbar$) should not be important, at least near 
150 GPa. Quantum corrections were therefore evaluated {\em a posteriori}
within the harmonic approximation (i.e. to first order in $\hbar$). 
To this aim, we generated molecular dynamics trajectories at very low 
temperature ($T \approx 2$ K), from which we extracted~\cite{kohanoff} 
the frequencies of the 21 modes (4 vibrons, 9 phonons and 8 
librons) corresponding to the $\Gamma$-point of the 4-molecule BZ. 
The ZPE, computed by summing over the above modes, was evaluated only
for structures $Pca2_1$ and $P2_1/c$, the closest competitors for phase II. 
In fact, a $Pa3$ space group is incompatible with experimental 
data~\cite{silvera2}, while all remaining structures are unstable in the 
clamped-nuclei approximation. The instability implies the existence of a 
distortion with negative clamped-nuclei energy curvature, and excludes 
stabilization through quantum corrections, to first order in $\hbar$. 
The ZPE turned out to be lower in $P2_1/c$ by only 0.003 eV/molecule 
with respect to $Pca2_1$, yielding a quantum-corrected enthalpy 
difference of 0.012 eV/molecule, still in favor of the classicaly 
determined $Pca2_1$ structure. 

It is worth stressing that in $Pca2_1$ the centers of the H$_2$
molecules occupy the sites of a slightly distorted {\it hcp} lattice. 
The temperature-induced
transformation of phase II into phase I should then be accompanied by
orientational disordering and by full recovery of the $hcp$ symmetry. 
Thus, the equation of state (EOS) of phases I and II are expected not 
to be significantly different. 
In Fig. 2 we report our EOS and the $c/a$ ratio calculated for the $Pca2_1$ 
structure in the range from 100 to 150 GPa. The $b/a$ ratio 
turns out to be essentially 
pressure-independent, although slightly smaller ($b/a\approx 1.715$) 
than the $hcp$ value of $\sqrt{3}$. 
Comparison with the experimental results for the EOS and the $c/a$ ratio 
in phase I~\cite{loubeyre} (full line) is also rather satisfactory. 

A structure with $Pca2_1$ symmetry is compatible with optical
measurements on phase II, where a strong Raman peak, and two IR modes are 
observed~\cite{silvera2,hemley1}. A symmetry analysis of the four vibronic 
modes of the $Pca2_1$ structure, usually labeled as $A_1$, $A_2$, $B_1$, 
and $B_2$~\cite{silvera2,turrell}, 
shows that all of them should be Raman active, and three of them 
($A_1$, $B_1$, and $B_2$) IR active. The frequencies of these four modes were 
evaluated by decomposing molecular dynamics trajectories at 100 and 150 GPa 
into the symmetry representations of space group $Pca2_1$~\cite{110}.
The resulting frequencies turned out to be substantially higher than 
experiment. Moreover, they {\em increased} as a function of 
pressure, in agreement with the calculated reduction of the 
molecular bond length (see Table I), but in flagrant disagreement with 
the observed trends, particularly for the Raman mode.

A closer glance at the experimental data, however, indicates that quantum 
effects are far from negligible (see Fig. 7 of Ref. \cite{hemley1}). 
In fact, the D$_2$ Raman frequency (scaled by $\sqrt{2}$) is roughly 200 
$cm^{-1}$ higher than that of H$_2$. To first order in $\hbar$, quantum 
corrections to the frequency are known to scale as the inverse of the 
particle's mass, no matter what the potential is~\cite{scheerboom}. Thus, 
using the experimental H$_2$ and D$_2$ frequencies, we obtain empirical 
but accurate quantum corrections to the classical values following
Ref.~\cite{ashcroft90}. In the case of the 
H$_2$ Raman mode the correction is as large as 500 cm$^{-1}$, while for the 
IR modes the effect is more modest, amounting to about 150 cm$^{-1}$. The 
resulting quantum-corrected vibron frequencies are reported in Fig. 3, and 
compared with experimental data on H$_2$\cite{hemley1}. The agreement with 
experiment is now very good, and shows that the Raman vibron turnover is 
a purely quantum effect. Moreover, the agreement allows us to interpret the 
experimental Raman mode as our $A_1$ mode: a totally symmetric, in-phase 
vibration of the four molecules in the unit cell. According to its symmetry 
classification, this mode should also be IR active. It does not appear in 
the experimental IR spectra, possibly due to a small oscillator strength. 
This is reasonable for a mode where all molecules oscillate in phase, 
since IR activity can only come from a small intramolecular asymmetry.

We attribute the two IR vibrons to the $B_1$ and $B_2$ modes. The predicted 
splitting of the two IR vibrons turns out to be in fairly good agreement 
with the data. Finally, we predict a Raman vibron, of symmetry $A_2$, with 
a ``classical'' frequency similar to that of the IR modes, being similarly an 
out-of-phase oscillation of the four molecules. Applying the same type of
quantum correction as for the IR modes, the $A_2$ should in principle be 
observed in the region around 4500 cm$^{-1}$. Failure to observe modes 
$A_2$, $B_1$ and $B_2$ in Raman spectroscopy could be attributed to their 
small Raman strength, contributions from out of phase vibrations canceling 
out. The observed Raman mode could alternatively be attributed 
to mode $A_2$, but this would imply a quantum-corrected frequency 
of 4370 cm$^{-1}$ (at 150 GPa), in clear disagreement with experiment.   
Symmetry lowering arguments and larger unit cells have been very recently 
invoked to explain a larger number of optical modes by band folding in 
ortho-para D$_2$ mixtures~\cite{goncharov}. The ortho-para distinction 
is however not included in our calculation.

Finally, our calculation provides the full electronic structure and 
pressure coefficients, and in principle the optical properties. 
In Fig. 4 we report the band structure and the electronic density of 
states of the $Pca2_1$ structure, at $P=100$ GPa (the density of states at 
150 GPa is also reported, for comparison). A gap of about 4 eV 
opens up the otherwise fairly free-electron-like bands. 
As shown (see also Table I), the gap decreases monotonically with 
pressure but is always finite and large in the range of stability of phase II. 

In summary, by means of extensive and accurate constant-pressure 
{\em ab-initio} molecular-dynamics simulations we have obtained 
what appears to be a definitive description of the broken symmetry phase 
of H$_2$. 
The $Pca2_1$ structure, first described by Kitaigorodskii and 
Mirskaya~\cite{KM} for the ground state of a classical quadrupolar system,
and later proposed for H$_2$~\cite{nagara}, is favored.
Equation of state and vibronic frequencies turn out to be
in very good agreement with available experimental data. 

This work was partly sponsored by INFM project LOTUS, and by the
European Commission under contract ERBCHRXCT 940438.

\begin{figure}
\caption{Candidates for the ground state structure of 
solid H$_2$ in the broken symmetry phase (phase II), projected along 
the $c$-axis. Black (gray) arrows represent molecules centered on the 
$c$ ($c$/2) plane and pointing towards the positive-$z$ hemisphere.}
\end{figure}

\begin{figure}
\caption{Calculated equation of state of H$_2$ in phase II:
8-atom simulation cell (solid circles), and
64-atom simulation cell (open squares). The experimental EOS of H$_2$ 
in phase I~[26]
is also reported (solid line). The inset shows the pressure dependence  
of the $c/a$ ratio (triangles are data from~[26], other
symbols as above).}
\end{figure}

\begin{figure}
\caption{IR and Raman vibron frequencies of H$_2$ as a function of 
pressure. The lower and the couple of upper solid lines are the experimental 
Raman and IR data~[3,5], 
respectively. Circles are calculated Raman vibrons 
(mode $A_1$). Squares and hexagons are calculated 
IR active vibrons (modes $B_1$ and $B_2$, respectively). Open triangles are 
estimates for the new $A_2$, Raman-only vibron.  }
\end{figure}

\begin{figure}
\caption{Electronic band structure (left panel) and density of states (right
panel) of solid H$_2$ in the $Pca2_1$ structure, at $P=100$ GPa. The energy
gap is 4.12 eV. The density of states at 150 GPa is also reported (dashed line,
right panel)} 
\end{figure}

\begin{table}
\caption{Bond length, volume, and band gap in the two
quadrupolar structures $Pca2_1$ and $P2_1/c$, as a function of pressure.}
 
\begin{tabular}{|c|c|c|c|c|}
 & $P$ (GPa) & $d_{HH}$~(\AA)& V~(cm$^3$/mole) &  $E_g$ (eV)\\ \hline
$Pca2_1$  &   100    &  1.377    &  2.75   &  4.12 \\
$Pca2_1$  &   150    &  1.373    &  2.35   &  2.53 \\ \hline
$P2_1/c$  &   150    &  1.374    &  2.35   &  2.61 \\
\end{tabular}
\end{table}


\begin{references}
\bibitem{wigner} E. Wigner and H. B. Huntington, J. Chem. Phys. {\bf 3},
764 (1935).
\bibitem{nellis} S. T. Weir, A. C. Mitchell, and W. J. Nellis, Phys. Rev.
Lett. {\bf 76}, 1860 (1996).
\bibitem{hemley1} H. K. Mao and R. J. Hemley, Rev. Mod. Phys. {\bf 66},
671 (1994) and references therein.
\bibitem{silvera1} I. F. Silvera in {\it Simple Molecular Systems at very 
High Pressure}, eds. A. Polian, P. Loubeyre and N. Boccara 
(Plenum, NY, 1989).
\bibitem{silvera2} L. Cui, N. H. Chen and I. F. Silvera, Phys. Rev. B
{\bf 51}, 14987 (1995).
\bibitem{barbee} T. W. Barbee III, A. Garcia, M. L. Cohen and J. L. Martins,
Phys. Rev. Lett. {\bf 62}, 1150 (1989), T. W. Barbee III and M. L. Cohen,
Phys. Rev. B {\bf 44}, 11563 (1991).
\bibitem{surh} M. P. Surh, T. W. Barbee III, and C. Mailhiot, Phys. Rev.
Lett {\bf 70}, 4090 (1993).
\bibitem{kaxiras} E. Kaxiras, J. Broughton and R. J. Hemley, Phys. Rev.
Lett. {\bf 67}, 1138 (1991); E. Kaxiras and J. Broughton, Europhys.
Lett. {\bf 17}, 151 (1992).
\bibitem{nagara} H. Nagara and T. Nakamura, Phys. Rev. Lett. {\bf 68},
2468 (1992).
\bibitem{mazin} I. I. Mazin and R. E. Cohen, Phys. Rev. B {\bf 52}, R8597
(1995).
\bibitem{hohl} D. Hohl et al., Phys.  Rev. Lett. {\bf 71}, 541 (1993).
\bibitem{klug} J. S. Tse and D. D. Klug, Nature {\bf 378}, 595 (1995).
\bibitem{focher} P. Focher et al., Europhys. Lett. {\bf 36}, 345 (1994).
\bibitem{scandolo} S. Scandolo and J. Kohanoff (unpublished)
\bibitem{beckper} A. D. Becke, Phys. Rev. A {\bf 38}, 3098 (1988); J. P.
Perdew and Y. Wang, Phys. Rev. B {\bf 33}, 8822 (1986).
\bibitem{giannozzi} P. Giannozzi (unpublished). See also F. Gygi, Phys.
Rev B {\bf 48}, 11692 (1993).
\bibitem{note} The calculated clamped-nuclei H$_2$ bond length and frequency 
are 0.752~\AA~and 4290 cm$^{-1}$, respectively. The exact value obtained 
within the Born-Oppenheimer approximation, i.e. corrected for 
anharmonic quantum effects, is 4050 cm$^{-1}$, to be compared with the
experimental gas-phase value of 4160 cm$^{-1}$.
\bibitem{ashcroft} B. Edwards, N. W. Ashcroft, and T. Lenosky, Europhys.
Lett. {\bf 34}, 519 (1996).
\bibitem{harrison} 
W.A. Harrison, {\it Solid State Theory} (McGraw-Hill, NY, 1970).
\bibitem{payne} I. J. Robertson and M. C. Payne, J. Phys.: Condens. Matter
{\bf 2}, 9837 (1990).
\bibitem{monkhorst} H. J. Monkhorst and J. D. Pack, Phys. Rev. B {\bf 13},
5188 (1976).
\bibitem{KM} A. I. Kitaigorodskii and K. V. Mirskaya, Sov. Phys.-- Cryst. 
{\bf 10}, 121 (1965).
\bibitem{natoli} V. Natoli, R. M. Martin and D. M. Ceperley, Phys. Rev.
Lett. {\bf 70}, 1952 (1995).
\bibitem{h16} This was done using a supercell contaning eight molecules such
that periodic images of the rotating molecule are not nearest neighbors.
\bibitem{kohanoff} J. Kohanoff, Comput. Mater. Science {\bf 2}, 221 (1994).
\bibitem{loubeyre} P. Loubeyre et al., Nature {\bf 383}, 702 (1996).
\bibitem{turrell} G. Turrell, {\em Infrared and Raman spectra of crystals}
(Academic Press, London, 1972).
\bibitem{110} A constant upward shift of 110 cm$^{-1}$ has been added to all
frequencies, so as to reproduce exactly the zero-pressure value \cite{note}.  
\bibitem{scheerboom} M. I. M. Scheerboom and J. A. Schouten, Phys. Rev.
B {\bf 53}, R14705 (1996).
\bibitem{ashcroft90} see expression (8) in: N. W. Ashcroft, Phys. Rev. B 
{\bf 41}, 10963 (1990). For the isolated molecule, we find that this
type of correction accounts for 90\% of the exact one.
\bibitem{goncharov} A. F. Goncharov et al., Phys. Rev. B. {\bf 54}, R15590
(1996).
\end{references}
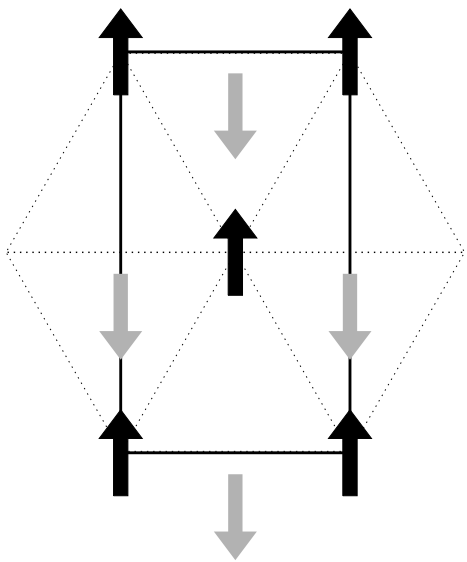
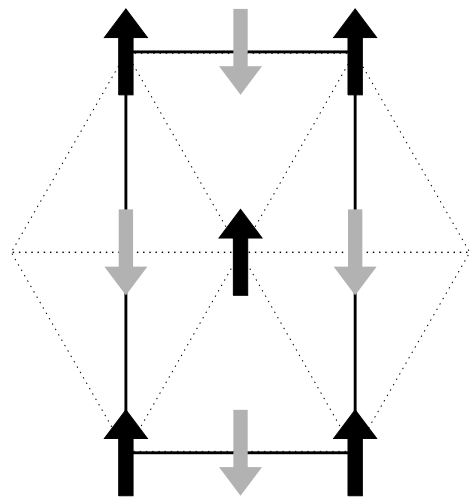
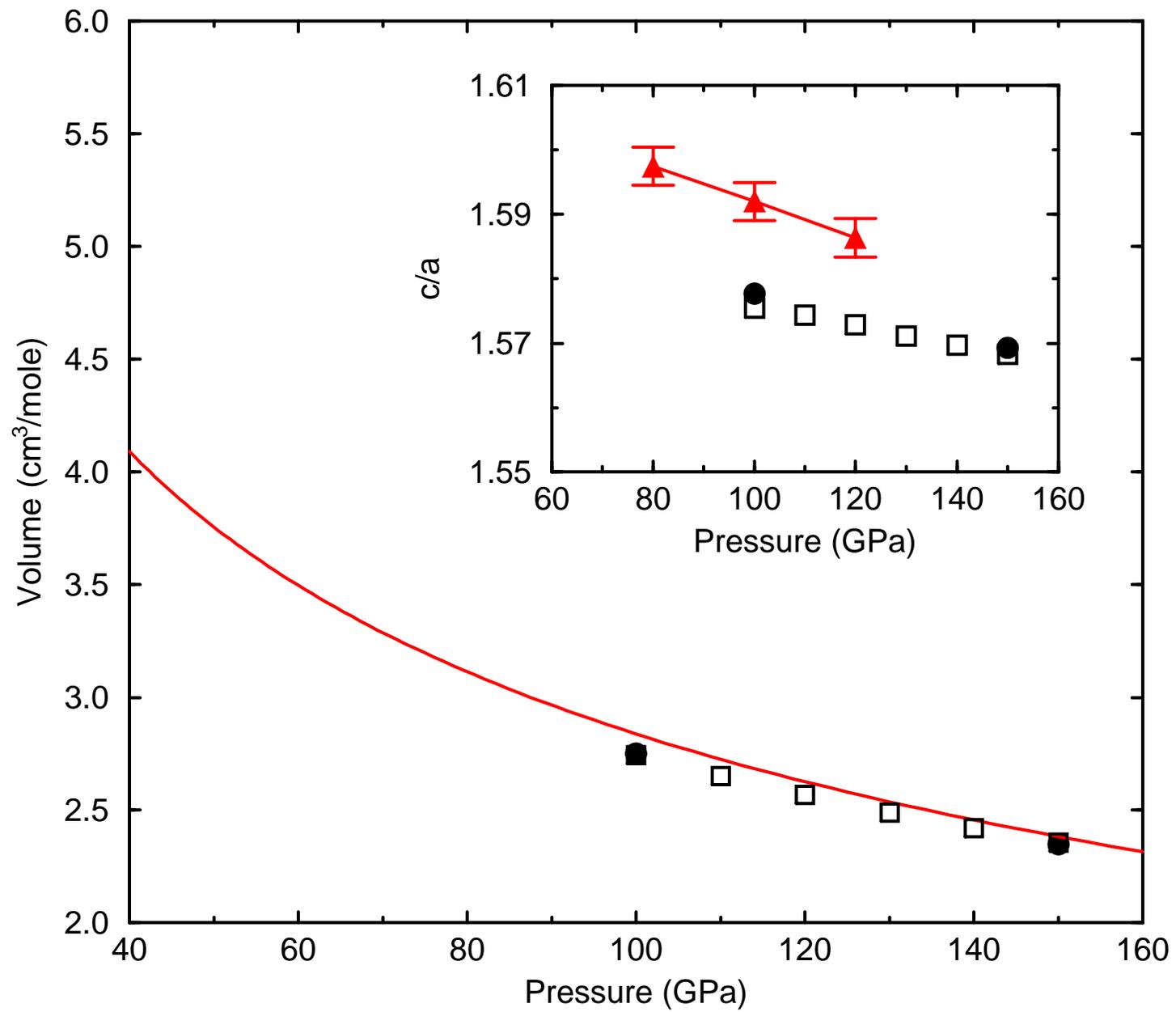
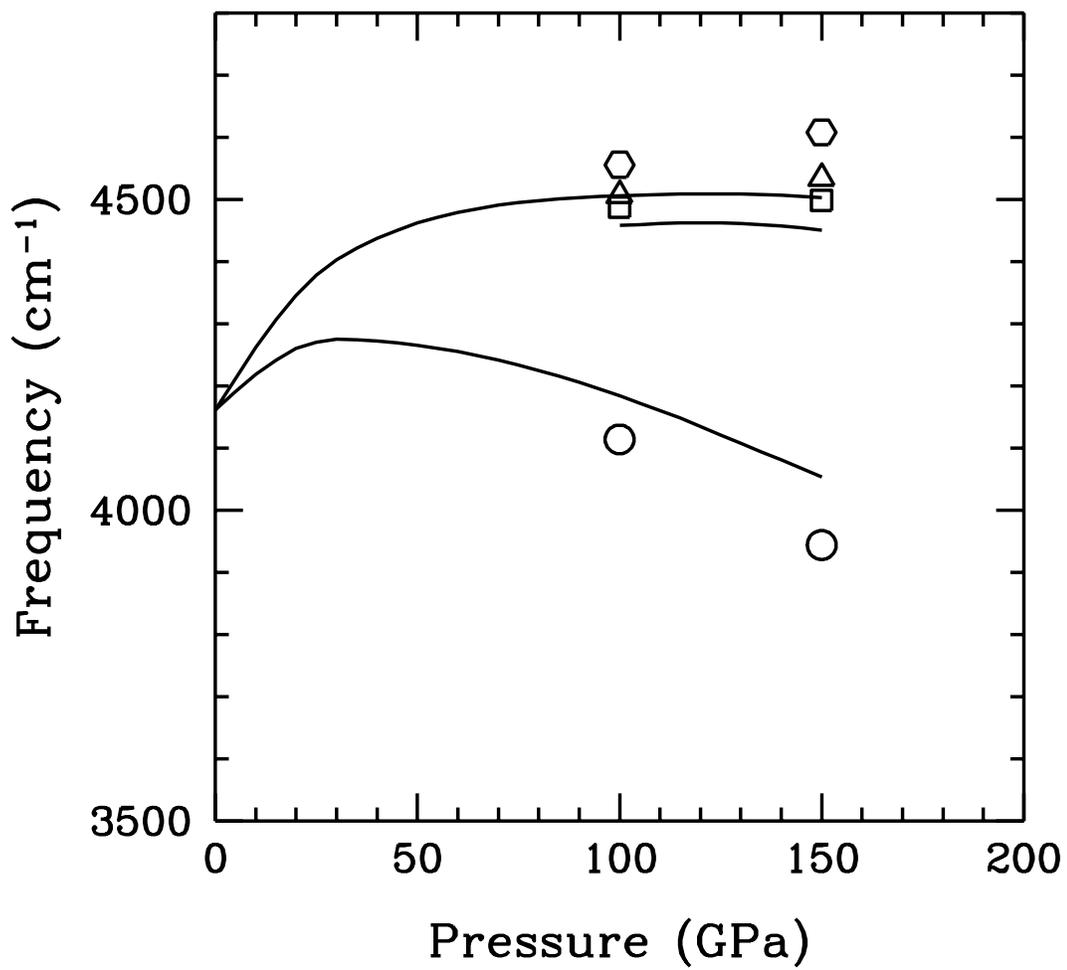
\end{document}